\documentclass[aps,prx,superscriptaddress,reprint]{revtex4-1}
 \UseRawInputEncoding
\usepackage[usenames,dvipsnames]{xcolor}
\usepackage{amsmath}
\usepackage{amssymb}
\usepackage{graphicx}
\usepackage{url}
\usepackage{hyperref}
\usepackage{color,xcolor}
\usepackage{epstopdf}
\usepackage{float}
\usepackage{ulem}
\usepackage[percent]{overpic}
\usepackage{siunitx}
\usepackage{subdepth}
\usepackage{multirow}
\usepackage{mathtools}

\usepackage{dcolumn}
\usepackage{bm}
\usepackage{times}
\usepackage[version=3]{mhchem} 

\usepackage{mathrsfs}
\usepackage{lipsum}
\usepackage{amsfonts}
\usepackage{array}
\usepackage{ulem}
\usepackage{lineno}
\usepackage[none]{hyphenat}

\hypersetup{colorlinks=true,citecolor=Red,linkcolor=Blue,urlcolor=Black}
\begin{document}
\title{Diabatic error and propagation of Majorana zero modes in interacting Quantum Dots systems}

\author{Bradraj Pandey}
\email{bradraj.pandey@gmail.com}
\affiliation{Department of Physics and Astronomy, The University of 
Tennessee, Knoxville, Tennessee 37996, USA}
\affiliation{Materials Science and Technology Division, Oak Ridge National 
Laboratory, Oak Ridge, Tennessee 37831, USA}

\author{Gaurav Kumar Gupta}
\email{ggupta3@cougarnet.uh.edu}
\affiliation{\mbox{Texas Center for Superconductivity and Department of Physics, University of Houston, Houston, Texas 77204, USA}}

\author{Gonzalo Alvarez}
\affiliation{Computational Sciences and Engineering Division, Oak Ridge National Laboratory, Oak Ridge, Tennessee 37831, USA}

\author{Satoshi Okamoto}                                                                               
\affiliation{Materials Science and Technology Division, Oak Ridge National                             
Laboratory, Oak Ridge, Tennessee 37831, USA}   

\author{Ronny Thomale}                                                                               
\affiliation{University of W\"urzburg, Am Hubland, 97074 W\"urzburg, Deutschland, Germany}   

\author{Elbio Dagotto}
\email{edagotto@utk.edu}
\affiliation{Department of Physics and Astronomy, The University of 
Tennessee, Knoxville, Tennessee 37996, USA}
\affiliation{Materials Science and Technology Division, Oak Ridge National 
Laboratory, Oak Ridge, Tennessee 37831, USA}

\date{\today}

\begin{abstract}
Motivated by recent experimental  progress in realizing Majorana zero modes (MZMs) 
using quantum dot systems, we investigate the diabatic errors associated with the movement of those MZMs. 
The movement is achieved by tuning time-dependent gate potentials applied to individual quantum dots, effectively creating a moving potential wall. In order to probe the optimized movement of MZMs, we calculate the experimentally accessible time-dependent fidelity and local density-of-states using many-body time-dependent numerical methods. Our analysis reveals that an optimal potential wall height is crucial to preserve the well-localized nature of the MZM during its movement. Moreover, for the first time, we analyze diabatic errors in realistic quantum-dot systems, incorporating the effects of repulsive Coulomb interactions and disorder in both hopping and pairing terms. Additionally, we provide a comparative study of diabatic errors arising from the simultaneous versus sequential tuning of multiple gates during the MZMs movement. Finally, we estimate the time scale required for MZM transfer in a six-quantum-dot system, demonstrating that MZM movement is feasible  and can be completed well within the qubit's operational lifetime  in practical quantum-dot setups.
\end{abstract}

\maketitle

\section{Introduction}
Majorana zero modes (MZMs) have potential applications in the development of fault-tolerant quantum computation~\cite{Kitaev1,Kitaev2,brien,Sarma,Nayak}. The non-local encoding of information and non-Abelian exchange statistics of MZMs allow for the construction of topologically protected qubits and gates, which are immune to local perturbations and decoherence. Experimentally achieving Majorana zero modes has been studied extensively through a hybrid semiconductor nanowire or a ferro-magnetic atomic chain system with strong spin-orbit coupling placed over a superconductor~\cite{Roman,Mourik}. However, in these systems the presence of disorder and impurities continues to pose significant challenges to the formation and detection of Majorana zero modes~\cite{Stanescu,Gomanko,Loo,PanH}.
Recently, a novel approach has been proposed to realize Majorana zero modes (MZMs) by utilizing a chain of quantum dots~\cite{Jay,Loss,Deng}. The experimental realization of the minimal Kitaev chain using just two quantum dots coupled via a short superconductor-semiconductor hybrid~\cite{Dvir}, has given new hope for the observation of Majorana zero modes. In these systems, the effective couplings between quantum dots are gate-tunable and mediated by well-controlled Andreev bound states in the hybrid semiconductor-superconductor region. The excellent local control over the chemical potential of each quantum dot significantly reduces the effects of disorder and smooth gate potentials compared to more conventional proximitized semiconductor nanowire systems ~\cite{Dvir, Bordin,ChunX,Zatelli}.

The conventional nano-wire SC heterostructures left several challenges open, such as to pin down the correct effective low-energy model. Meanwhile, the microscopic modeling of the new quantum-dot platform is tentatively simpler and more closely tied to the original Kitaev toy model. Very recently, using three quantum dots, and working at the ``sweet spot"  $t_h=\Delta$ (where the electronic hopping $t_h$ and superconducting coupling $\Delta$ are equal in magnitude)~\cite{ChunX} two peaks in tunneling conductance measurements were observed interpreted as localized MZMs.  The topological protection of MZM increased compared to the Kitaev experiments with two sites~\cite{Bordin,ChunX}.  In the quantum-dot system the MZMs are perfectly localized onsite at the ``sweet spot"  $t_h=\Delta$ . Remarkably, an enhancement of the excitation gap using Yu-Shiba-Rusinov states was shown experimentally in quantum-dot systems~\cite{Zatelli}.  These quantum-dot systems  offer a promising platform for quantum information processing with Majorana zero modes (MZMs)~\cite{Liu,Mazur}, where the manipulation of MZMs can be precisely controlled using local gates. The controlled movement of MZMs across three quantum dots has been experimentally realized recently by tuning the chemical potential of the outer dots~\cite{Bas}. 
 
 The optimal transport of Majorana zero modes (MZMs) is crucial for braiding or fusion of MZMs in the broad framework of topological quantum computation~\cite{Alicea,Aasen,Rachel1}. However, an important issue is that the manipulation of MZMs in quantum-dot systems for such applications {\it must be completed within the quasiparticle ``poisoning" time of the MZMs}~\cite{ChunX}. Specifically, a rapid movement of MZMs can induce quasiparticle excitations between the ground-state manifold and the excited states, resulting in the loss of quantum information. Therefore, it is essential to manipulate MZMs within a finite window in time that (i) makes quantum computation doable, while simultaneously (ii) minimizing unwanted diabatic transitions~\cite{MENG,perfetto,LXU}.

For nanowire systems, the optimized transport of Majorana zero modes (MZMs) has been proposed in several studies, primarily using a single-particle formalism~\cite{Oppen,Mishmash,Conlon,Coopmans}. {{In arrays of tunable gates, the diabatic error during MZM movement has been explained using a simple Landau-Zener description~\cite{Mishmash}. For MZM transport using a single moving potential wall, it has been shown that a critical velocity, $v_{\text{crit}} = \Delta_{\text{eff}}$ (the effective gap), exists above which the moving-frame Hamiltonian becomes gapless, resulting in the loss of topological protection for MZMs~\cite{Karzig,Conlon}. Recently, in non-interacting nanowire systems, it was demonstrated that an optimal number of gates is required to move MZMs effectively while minimizing diabatic error~\cite{Truong}. However, the diabatic error in interacting systems remains largely unexplored~\cite{Sekania,Bpandey1}, despite its significant role and impact on MZM-based quantum computing.}}

\begin{figure}
\centering
\begin{overpic}[width=1.0\columnwidth]{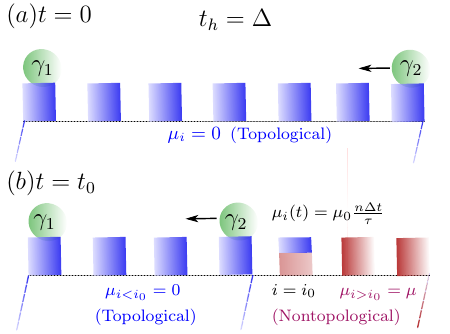}
\end{overpic}
  \caption{Schematic representation of the movement of the right edge MZM $\gamma_2$ in a coupled quantum-dots system using individually 
    tunable gates attached to each quantum dot. (a) At time $t=0$, the system has two MZMs ($\gamma_1$, $\gamma_2$) 
    located on the left and right edges of the system at the sweet-spot $t_h=\Delta$.
    (b) At time $t=t_0$, the right edge MZMs moved to site $i=i_0$ by using a time-dependent onsite chemical potential $\mu(t)$ (($\mu_0$ is the height of the potential wall)).
    The onsite chemical potential can be tuned by application of local gates attached to quantum-dots with quench rate $\tau$.
    The blue color regions are  topological while the red color regions are non-topological}
\label{fig1}
\end{figure}

In this work, we analyze the diabatic errors that arise during the movement of MZMs in an interacting quantum-dot system, employing a {\it many-body formalism}. Compared to nanowires, in quantum-dot systems at the sweet spot, MZMs are localized on single sites of the left and right edges of the Kitaev chain~\cite{Dvir,ChunX}. To enable the movement of MZMs, we assume that each quantum dot is connected to a time-dependent gate, which modulates the local chemical potential of the individual quantum dot~\cite{Alicea}. This modulation creates a moving potential wall (see Fig.~\ref{fig1}) by transitioning a topological quantum dot (blue) to a non-topological quantum dot (red). 

To study diabatic errors during the movement of MZMs in the quantum-dot system, we calculate the time-dependent {\it fidelity}~\cite{Truong} as well as the time-dependent local density-of-states (LDOS) for each quantum dot~\cite{Bpandey1,Bpandey2}. First, we focus on the height of the potential wall required to move the localized MZM from one quantum dot to another. We find that optimizing the wall height is crucial to maintaining the well-localized behavior of the MZM, and measuring the local density-of-states plays a vital role in this optimization. Second, we analyze the effects of the nearest-neighbor Coulomb interaction on diabatic errors during the movement of MZMs. In realistic quantum dot systems the hopping and pairing terms may not be perfectly equal for a moderate number of quantum dots, leading to disorder in these terms~\cite{ChunX,Jay}. For this reason, we investigate the diabatic error in the presence of disorder in the hopping and pairing terms and find that the diabatic error increases very slowly for moderate disorder. However, a further increase in disorder strength leads to a jump in the diabatic error. Third, we investigate the diabatic error during MZM movement in quantum-dot systems for the case when multiple gates are tuned simultaneously, using both linear and sine-squared ramping protocols. We find that the sine-squared ramping protocol is much more efficient and produces significantly less error than the linear ramp protocol. Our results show that for slower tuning rates of the gate potentials, the diabatic error is significantly reduced in the case of multiple-gate tuning compared to single-gate tuning. Finally, using  six quantum-dots, we estimate the time scale to move the MZMs in realistic quantum-dot experimental setups~\cite{ChunX}.  

{{The organization of the manuscript is as follows. Section \ref{sec:model} introduces the interacting Kitaev model for quantum-dot systems at the sweet spot and provides details about our numerical simulations. Section \ref{sec:results} presents the results of our study for the movement of MZMs, including the effect of the wall height (Sec. \ref{sec:results:wall}), nearest-neighbor Coulomb repulsion (Sec. \ref{sec:results:Coulomb}), and disorder in the hopping and pairing terms (Sec. \ref{sec:results:disorder}). Section \ref{sec:results:gate} compares the diabatic error during the movement of MZMs for simultaneous and sequential gate tuning. Finally, Section \ref{sec:conclusion} provides our conclusions and discussion. In the Appendices, we include results on the non-analytic behavior of the many-body gap (Appendix \ref{appendix:gap}) and the diabatic error during MZM movement in the presence of disorder in the on-site potential (Appendix \ref{appendix:onsite}).}}

\section{Model and Methods} \label{sec:model}
\begin{figure}
\centering
\begin{overpic}[width=1.0\columnwidth]{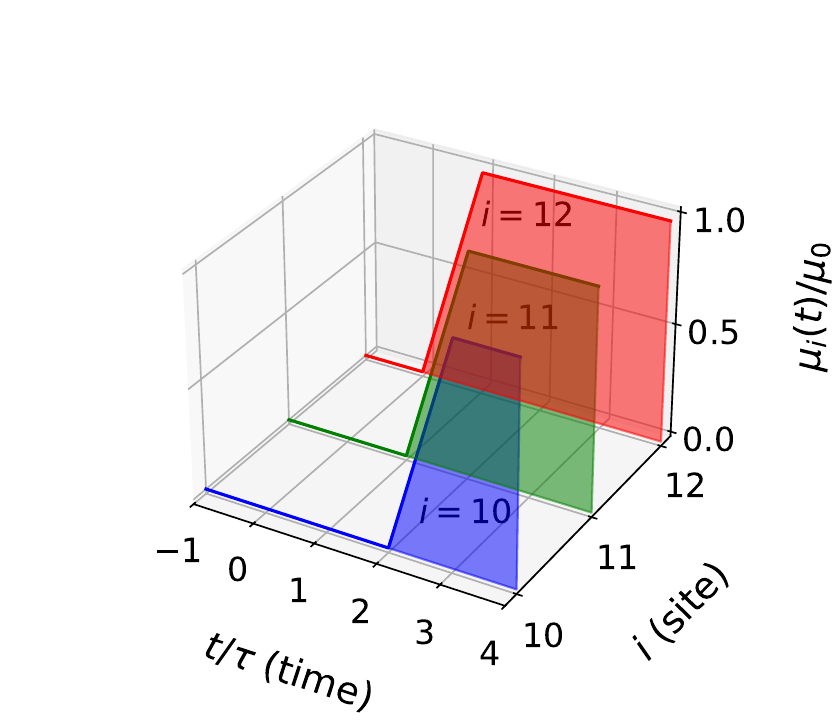}
\end{overpic}
\caption{Ramping protocol to move a MZM through the system. At time $t/\tau=0$, we prepare the system in the ground state with all the onsite potential set at $0$. For a time interval $0\leq t/\tau\leq1$, the onsite potential at the right most site (site $i=12$) is increased from $0$ to $\mu_0$ {\it linearly} and stays constant on-wards. Later on, in the time interval $1\leq t/\tau \leq 2$, the onsite potential at the rightmost site remains constant, but the onsite potential at the next site ($i=11$) increases linearly until it reaches $\mu_0$.}
\label{fig2}
\end{figure}

The recent progresses in quantum-dot systems have enabled the experimental realization of two- and three-sites Kitaev chains~\cite{Dvir,ChunX}. In these setups, spin-polarized quantum dots and  short superconductor-semiconductor hybrids  are coupled by elastic co-tunnelling (ECT) and crossed Andreev reflection (CAR)~\cite{Dvir,ChunX}. This configuration allows for tunable single-electron hopping $t$ (ECT) and triplet pairing $\Delta$ (CAR), both mediated by Andreev bound states in the hybrid segments~\cite{Liu}. These couplings  in the quantum-dots systems are tuned by  electrostatic gates connected to the SC-SM hybrid segment and  provides all the  key ingredients for implementing artificial Kitaev chains~\cite{Dvir}. The Kitaev chain Hamiltonian for spinless fermions, with hopping $t$ and $p$-wave pairing $\Delta$, can be described as~\cite{Kitaev1}:
\begin{eqnarray}
    H_0 &=& -t_h\sum_{i=1}^{N-1}\left( c_i^\dagger c_{i+1}^{} + \text{H.c.}\right) + V\sum_{i=1}^{N-1} n_i n_{i+1} \nonumber\\
    && \Delta\sum_{i=1}^{N-1}\left( c_i c_{i+1} + \text{H.c.} \right),
    \label{eq_ham}
\end{eqnarray}
where $c_i (c_i^\dagger)$ is the fermionic anhilation (creation) operator, $n_i = c_i^\dagger c_i$ is the number operator, and $t_h$ and $\Delta$  are the nearest-neighbour hopping and SC pairing amplitudes, respectively.  Throughout the calculation, we set $t_h = 1$ and $\Delta = t_h$ (unless stated otherwise), a condition that has been realized in quantum-dot systems~\cite{Dvir}. In the case of a two-site quantum-dot system, two perfectly localized MZMs were observed at the sweet spot $2t_h = 2\Delta = 25 \,\mu eV$. The tuning of the coupling ratio $t_h/\Delta = 1$ is easily achieved by adjusting the electrostatic gate potential in the quantum-dot system, which enforces the product-state property of the ground state and thereby yields maximally localized Majorana states~\cite{Thomale2}. Using the Hamiltonian in Eq. \ref{eq_ham}, we calculate the ground state using exact diagonalization (ED)~\cite{Bpandey1} and/or state-of-art density matrix renormalization group (DMRG)~\cite{Alvarez1,Nocera}. At time $t=0$, for the parameters $t_h=\Delta=1$, and $V=0$, we have two onsite localized MZMs ($\gamma_1$ and $\gamma_2$) on the left and right edges of the Kitaev chain (see Fig.~\ref{fig1}(a)). The right edge MZM ($\gamma_2$) is moved using sequential operations of the time-dependent gate potential (Fig.~\ref{fig1}(b)).  We introduce the time dependence by adding a time-dependent chemical potential in the Hamiltonian as follows~\cite{Bpandey1}

\begin{eqnarray}
    H(t) &=& H_0 + \sum_i \mu_i(t) n_i\\
    \mu_i(t) &=&
\begin{dcases}
    0,              & i<i_0\\
    \mu_0\frac{t}{\tau},& \text{if } i= i_0\\
    \mu_0,            & i > i_0
\end{dcases}
\label{eq_t_ham}
\end{eqnarray}
where $1/\tau$ is the quench rate. We discretize the time steps as $t = k\Delta t$, where $\Delta t = 0.001$ is a small time step,  
and $k$ is an integer chosen such that the chemical potential at site $i = i_0$ varies nearly linearly  from $0$ to $\mu_0$ over a time $\tau$. We first choose a linear time-dependent chemical potential since, in quantum-dot experiments for the two-site Kitaev model, the chemical potentials of quantum dots are linearly tuned by voltages applied via gates~\cite{Dvir}. We also compare the sine-squared ramping protocol with the linear ramping protocol.  
 The time-dependent onsite chemical potential  
can be visualized as a moving potential wall that grows at site $i = i_0$ over time $\tau$,  
then shifts to the next site $i = i_0 - 1$, and grows again, as illustrated in Fig.~\ref{fig2}.
  This process is analogous to gate switching in real experimental setups. As a result, the MZM on the right edge moves toward the center, while the MZM on the left edge remains localized (Fig.~\ref{fig1}).

The ground state, $|\psi(0)\rangle$, of $H_0$ in  Eq. (\ref{eq_ham})  was calculated using ED/DMRG and then evolved with the time dependent Hamiltonian $H(t)$ in Eq. (\ref{eq_t_ham}) using the time-evolution operator, $U(t) = \mathcal{T} \text{exp}{\left (-i\int_{0}^{t}H(s) ds\right)}$, where $\mathcal{T}$ is the time ordering operator~\cite{Slingerland}. Numerically, the time-evolution was performed using time-dependent ED and the time-dependent variational principle (TDVP) using the  ITensor Library~\cite{Fishman}. One-dimensional chains of varying lengths up to $L = 36$ were studied and the bond dimension for DMRG calculations were kept up to $m = 120$ which ensures an accuracy involving a truncation error around $1.0\times 10^{-6}$. The fidelity was then calculated by calculating the overlap of  the ground state wave-function of the instantaneous Hamiltonian $H_I$ with the time evolved wave-function i.e, $\mathcal{F}(t) = |\langle \psi_g(I) |\psi(t)\rangle| = |\langle \psi_g(I)| U(t)|\psi(0)\rangle|$. We then calculate the diabatic error via  $f(t) = 1 - \mathcal{F}(t)$.

 In addition to fidelity, we also calculate the time-dependent local density-of-states (at zero temperature).  As described in Ref.~\cite{Bpandey1}, first we time evolve the initial state $|\psi(0)\rangle$  up to time $t=N_R\tau$, (here $N_R$ is the number of sites where the chemical potential reaches its maximum value $\mu_0$) using the time-dependent Hamiltonian $H(t)$ as: $|\Psi(t)\rangle= U(t)|\psi(0)\rangle$. In order to obtain the time and energy resolved local density-of-states, we calculate the double-time Green function $G(t,t')$~\cite{Kennes},  using the instantaneous Hamiltonian $H_f = H(t=t_f)$ at time $t_f=N_R \tau$:                                 
\begin{equation}                                                                                        G^{elec}_j(t,t')= \langle \Psi(t)|c^{\dagger}_j e^{iH_ft'}c_j e^{-iH_ft'}|\Psi(t)\rangle.              \end{equation}                                                                                         The time-dependent energy resolved $LDOS_{elec}$($\omega,j,t$) for electrons is given by $LDOS_{elec}(\omega,j,t)= \frac{1}{\pi} $Im$ \int_0^{T} dt' e^{i \left(\omega+i \eta\right) t'} iG^{elec}_j(t,t')$. In practice, we use $T=60$ for the integration with respect to $t'$, and a broadening $\eta=0.1$. 
Similarly as described in Ref.~\cite{Bpandey1}, we also calculate the hole portion of the local density-of-states $LDOS_{hole}$($\omega,j,t$). The time-dependent total local density-of-states at site $j$ can be obtained as $LDOS(\omega,j,t) =LDOS_{hole}(\omega,j,t)+LDOS_{elec}(\omega,j,t)$.

\section{Results}\label{sec:results}
In this section, we present the results for the movement of Majorana zero modes (MZMs) in quantum-dot systems, focusing on the effects of the wall height, the strength of the nearest-neighbor Coulomb interactions, and the effect of disorder in the hopping and pairing terms. Furthermore, we compare the diabatic error during the MZM movement when multiple gates are tuned simultaneously.

\subsection{Effect of  wall height during the movement of MZMs}\label{sec:results:wall}

\begin{figure*}[!ht]
\hspace*{-0.5cm}
\vspace*{0cm}
\begin{overpic}[width=2.0\columnwidth]{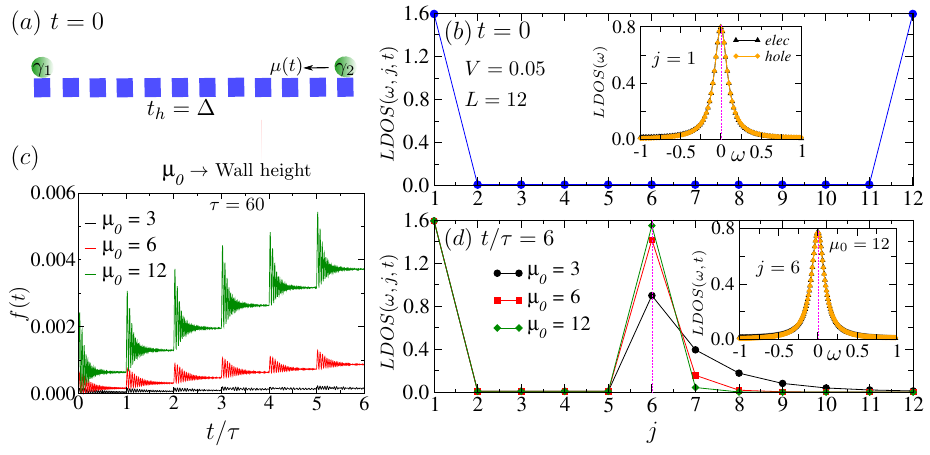}
\end{overpic}
\caption{(a) Schematic representation of left and right MZMs ($\gamma_1$, $\gamma_2$)  at time $t=0$
  and for $L=12$ quantum-dots at the sweet-spot $t_h=\Delta$. (b) Site-dependent local density-of-states $LDOS(\omega,j,t)$ vs site $j$,
  at $t=0$, $\omega=0$, and $V=0.05$. The sharp peaks at $j=1$ and $12$ represent localized MZMs at the edge sites. The inset shows
  the local density-of-states $LDOS(\omega)$ vs $\omega$ at $t=0$ for site $j=1$. (c) The diabatic error $f(t)$ as function of 
  $t/\tau$ for different values of $\mu_0$ and for $\tau=60$. (d) The site-dependent local density-of-states $LDOS(\omega,j,t)$ vs site $j$
  at time $t/\tau=6$ for different values of $\mu_0$. For larger values of $\mu_0=12$, the MZM $\gamma_2$ is sharply localized at 
   $j=6$. For lower values of $\mu_0$, the MZM $\gamma_2$ delocalizes over a few sites, which is not a suitable case to perform real quantum computation
   or braiding processes. The inset shows the $LDOS(\omega,t)$  at site $j=6$ for $\mu_0=12$ and $\tau=60$ 
  at $t/\tau=6$. 
}
\label{fig3}
\end{figure*}

 In this subsection, we analyze the effect of wall height on the movement of the right edge MZM ($\gamma_2$), by varying the constant term $\mu_0$ in Eq.~3.  {{To move the MZM from one site to another within a finite time $t$, the onsite chemical potential, $\mu_i(t)$, must be larger than the critical value $\mu_c > 2t_h$~\cite{Alicea},  in order to switch a site from a topologically nontrivial phase to a trivial phase. To track the movement of the MZM, we calculate both the time-dependent fidelity and the local density of states.}} At $t=0$, we choose parameters $t_h=\Delta=1$, $L=12$, and $V=0.05$, resulting in two localized MZMs $(\gamma_1,\gamma_2)$ at the left and right edges of the chain, as shown in Fig.~\ref{fig3}(a). The site-dependent local density-of-states, $LDOS(\omega,j,t)$, at $t=0$,  exhibits sharp peaks at sites $j=1$ and $12$, indicating two localized MZMs on the left and right edges of the Kitaev chain (Fig.~\ref{fig3}(b)). The inset of Fig.~\ref{fig3}(b) shows the electron and hole part of the local density-of-states $LDOS(\omega)$ as a function of $\omega$ for the right edge site $j=12$ and at $t=0$. The equal-height sharp peaks of the electron and hole components of $LDOS(\omega)$ at $\omega=0$ confirm the presence of a localized $\gamma_2$, on the right edge of the chain ~\cite{Bpandey3,Bpandey4,Herbrych}.

As described in the previous section, the right-edge Majorana zero mode (MZM), $\gamma_2$, can be moved via the sequential modulation of the onsite chemical potential, $\mu_i(t)$. In Fig.~\ref{fig3}, we examine the diabatic error, $f(t)$, for three different cases where $\gamma_2$ is moved using wall heights with $\mu_0 = 3$, $6$, and $12$. The MZM at the right edge is transported from site $j = 12$ to $j = 6$ in the same total time $T$ for the three cases. As shown in Fig.~\ref{fig3}(c), the diabatic error is the smallest for $\mu_0 = 3$ and increases with higher wall heights. This behavior arises because the construction of higher walls requires a faster rate of change in $\mu_i(t)$ compared to lower walls (for the cases when the total ramp time $\tau$ is kept fixed). The sudden change in chemical potential at the start of wall formation at each site $j$ induces oscillations in the diabatic error, $f(t)$. For faster changes in $\mu_i(t)$, these oscillations exhibit larger amplitudes due to the more pronounced nonanalytic behavior associated with a rapid modulation of the chemical potential~\cite{Truong,Garraway,Vitanov}. Furthermore, we observe that the effective gap (defined as the energy difference between the ground state and the first excited state) shows discontinuities only at the beginning of increasing the on-site chemical potential at each site (see Fig.~\ref{fig7} in the Appendix \ref{appendix:gap}).

In Fig.~\ref{fig3}(d), we present the real-space, time-dependent local density-of-states, $LDOS(\omega, j, t)$, for three different wall heights, $\mu_0 = 3$, $6$, and $12$, at time $t/\tau = 6$. Interestingly, for $\mu_0 = 3$, the right-edge MZM, $\gamma_2$, is not well localized near the wall at site $j = 6$, even though the diabatic error is smallest in this case. As $\mu_0$ increases from $6$ to $12$, the right-edge MZM, $\gamma_2$, becomes increasingly localized at site $j = 6$ (see Fig.~\ref{fig3}(d). A well-localized MZM is crucial for the braiding of MZMs, as it reduces the possibility of overlap with other nearby MZMs. In the inset of Fig.~\ref{fig3}(d), we show the time-dependent electron and hole components of the local density-of-states, $LDOS(\omega, t)$, as a function of $\omega$ at $t/\tau = 6$. The equal-height peaks of the electron and hole components of $LDOS(\omega, t)$ at $\omega = 0$ confirm the localization of $\gamma_2$ at site $j = 6$ at time $t/\tau = 6$. 

The main message of this subsection is that in addition to fidelity, it is crucial to measure the local density-of-states to ensure the accurate movement of localized MZMs. {{Namely, fidelity alone does {\it not} provide the optimal conditions for MZM movement, as the diabatic error remains minimal for slower movement irrespective of the wall height. In contrast, the site- and time-dependent local density-of-states offers a clearer understanding of the localization or delocalization of MZMs during their transport.}}

\subsection{Effect of  nearest-neighbor Coulomb interactions}\label{sec:results:Coulomb}

In this subsection, we explore the effect of repulsive nearest-neighbor Coulomb interactions, $V$, on the movement of the right-edge MZM, $\gamma_2$, in a moderately sized quantum-dot array with $L = 16$. For very small quantum dot systems, the nearest-neighbor Coulomb interactions are highly detrimental to the stability of MZMs, as the finite $V$  leads to an overlap between the MZMs~\cite{Bpandey2}. 
In realistic scenarios, the Coulomb interactions between neighboring quantum dots should be always present. However, its influence in moderately sized quantum dot chains is not significant.  This is because the nearest-neighbor Coulomb interactions, $V$, can be screened and minimized by the central hybrid superconducting region situated between neighboring quantum dots~\cite{Tsintzis}.

\begin{figure}
\centering
\begin{overpic}[width=1.0\columnwidth]{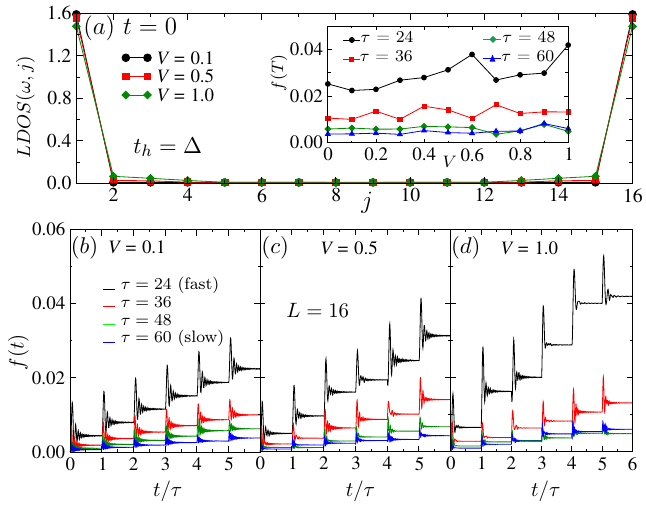}
\end{overpic}
  \caption{(a) Site-dependent local density-of-states $LDOS(\omega,j)$ vs $j$ at $t=0$ and for different values of the Coulomb interaction $V$
  using a system size $L=16$.
  For smaller values of $V\leq 0.1$, the MZMs are localized sharply on the edges. With an increase in $V$, the MZMs are localized exponentially, namely spread
  over a few more sites. The inset shows the $LDOS(\omega)$ vs $\omega$ for site $j=16$ and different values of  $V$.
  The diabatic error $f(t)$ vs $t/\tau$ for different values of $\tau$ and for Coulomb interactions as shown in (b) $V=0.1$, (c) $V=0.5$, and (d) $V=1.0$.
  The diabatic error increases with increase in $V$ and for smaller values of $\tau$, namely for fast movement, as intuition indicates. 
  } 
\label{fig4}
\end{figure}

Figure~\ref{fig4}(a) shows the site-dependent local density-of-states, $LDOS(\omega, j, t)$, at $t = 0$, for different values of the Coulomb interaction, $V$, using a system size of $L = 16$. For smaller values of $V$, the $LDOS(\omega = 0, j, t = 0)$ exhibits sharply localized onsite peaks at the edge sites $j = 1$ and $j = 16$. However, as $V$ increases, the local density-of-states becomes slightly delocalized over multiple sites. In the inset of Fig.~\ref{fig4}(a), we plot the diabatic error $f(T)$ at the final time $T$ (when the right edge MZM $\gamma_2$ reaches site $j=10$). In the lower panel of Fig.~\ref{fig4}, we plot the time-dependent diabatic error, $f(t)$, for different values of $V$ while moving the right-edge MZM, $\gamma_2$, from site $j = 16$ to site $j = 10$. For faster movement of $\gamma_2$ (corresponding to smaller values of $\tau = 24$), the diabatic error increases more rapidly with increasing repulsive Coulomb interaction, $V$. This occurs because, as $V$ increases, the effective energy gap (the difference between the ground state and the first excited state) decreases~\cite{Stoudenmire,Gangadharaiah}, leading to stronger interaction-induced non-adiabatic transitions. Consequently, the faster modulation of $\mu_i(t)$ causes the time-evolving state, $|\psi(t)\rangle$, to overlap with multiple higher-energy states~\cite{Bpandey1}. This overlap with higher-energy states results in an increased diabatic error for smaller $\tau$ values  with an increase in $V$ [see the inset of Fig.~\ref{fig4}(a), and Figs.~\ref{fig4}(b,c,d)].

For $\tau > 36$, we observe that the diabatic error decreases significantly compared to the faster changes in the gate potential, $\mu_i(t)$. Interestingly, for the slow movement of the right-edge MZM, $\gamma_2$, the diabatic error $f(T)$ does not vary significantly with increasing $V$ up to moderate values of the Coulomb repulsion (see the inset of Fig.~\ref{fig4}(a)). This behavior arises because the slower movement of the MZM generates fewer quasi-particle excitations above the reduced effective gap. We find that the time-evolved state, $|\psi(t)\rangle$, exhibits a very small overlap with higher-energy excited states, resulting in a very small diabatic error, $f(t)$, even as time progresses for moderate values of $V$ (see the inset of Fig.~\ref{fig4}(a) and the full Figs.~\ref{fig4}(b,c,d)).  
These results show that in the presence of a repulsive Coulomb interaction $V$, we need to move the MZMs using a quite slow rate in order to minimize the diabatic error. In the presence of Coulomb repulsion, the effective gap decreases, and a faster rate of change induces more quasi-particle excitations, leading to the loss of quantum information.

\subsection{Effect of disorder in hopping and pairing terms}\label{sec:results:disorder}

Experimental systems are inevitably prone to disorder, which may add additional features to the pristine system studied in the previous sections. To understand the effect of disorder on the transport of MZMs, we here add a site-dependent disorder component in the hopping amplitude ($t_h^k$) as well as in the SC pairing amplitude ($\Delta_k$) ~\cite{Jay} drawn from a normal distribution given by
\begin{eqnarray}
    g(x)  = \frac{1}{\sigma \sqrt{2\pi}}e^{-\frac{1}{2}\left(\frac{x-a}{\sigma}\right)^2}
\end{eqnarray}
where $\sigma$ is the standard deviation and $a$ is the mean of the distribution.  {{We assume that the onsite chemical potential at time $t=0$ is zero for each site, achievable in quantum dot experiments by tuning the gate potentials attached to individual quantum dots~\cite{ChunX}.}} For the calculations, we choose the distribution to be centered around zero ($a = 0$) with different standard deviations $\sigma$ up to $0.6$. 

{{ For the static case, the results in Ref.~\cite{Jay} demonstrate that even a small number of quantum dots is sufficient to support a robust pair of localized MZMs under moderate disorder, in contrast to nanowire systems.}}  We then study the effect of disorder on the movement of the right-edge MZM, $\gamma_2$, for a moderately sized quantum dot array with $L = 16$.  

\begin{figure}[t]
\centering
\includegraphics[width=0.48\textwidth]{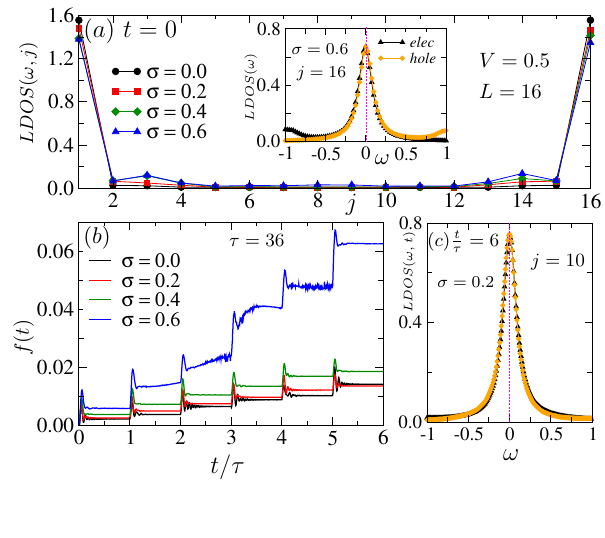}
\caption{(a) Site-dependent disorder averaged local density-of-states $LDOS(\omega,j)$ vs $j$ at time $t=0$  for different 
  values of disorder strength $\sigma$ introduced in the hoppings $t^h_k$ and in the pairing terms $\Delta_k$. Inset shows the 
$LDOS(\omega,j)$ vs $\omega$  at site $j=16$ and for $\sigma=0.6$. The equal-height peaks of the electron and hole parts of the $LDOS(\omega)$
  illustrate the existence of MZMs in the system for  $\sigma \leq 0.6$ at $t=0$. (b) The disorder averaged diabatic error $f(t)$ in the system during movement of the 
  right MZM $\gamma_2$ from site $j=16 \to 10$, for different values of $\sigma$. Clearly, by increasing $\sigma$ the diabatic
  error slowly increases with time $t/\tau$. For $\sigma =0.6$ the diabatic error takes a larger jump and the 
system may have low-quality MZMs. (c) Electron and hole parts of the time-dependent $LDOS(\omega,j,t)$ 
  at site $j=10$ for $\sigma=0.2$ for a particular value of the disorder realization (taken from a Gaussian distribution). 
}
\label{fig5}
\end{figure}

Figure \ref{fig5}(a) shows the site-dependent disorder averaged (using 40 realizations of disorder) 
$LDOS(\omega, j, t)$ at $t=0$ for varying $\sigma$ up to $0.6$ and nearest-neighbour interaction $V = 0.5$. We can observe that the MZMs are still localized at the edges of the quantum-dot array but have developed a tail with finite DOS, which increases as $\sigma$ increases. This is an expected behavior because MZMs are exponentially localized, with the shortest localization length for the case $t_k^h/\Delta_k = 1$. As disorder steers $t_k^h / \Delta_k$ away from unity, the localization length increases, and we start to observe the exponential tail in the $LDOS$. The inset of Fig. \ref{fig5}(a) shows the $LDOS$  as a function of $\omega$ for electron and hole at $j = 16$ and $\sigma = 0.6$, at $t =0$. Again, the equal-height sharped peaks of the electron and hole portions at $\omega = 0$ confirms the presence of the localized MZM $\gamma_2$. Additionally, we obtain a finite $LDOS$ for $\omega$ close to the SC coherence peak for the electron and hole part. This is due to the fact that the disorder introduces localized states near the SC coherence peaks, which extends into the SC gap.  

Figure~\ref{fig5}(b) shows the disorder-average diabatic error $f(t)$ acquired by the system during the movement of the MZM $\gamma_2$ from site $j=16$ to site $j=10$ for different values of $\sigma$ up to $0.6$ and a fixed $\tau = 36$.  We find that the diabatic error slowly increases with time for all values of $\sigma$, but for weak disorder, such as $\sigma = 0.2$, $f(t)$ is very close to the pristine system ($\sigma = 0.0)$. However, for stronger disorder, such as $\sigma=0.6$, we found a sudden increase in the diabatic error. This is due to the fact that a strong deviation of $t_h^k/\Delta_k$ from $1$ reduces the many-body gap and makes it easier to have low-energy excitations as we try to move the MZM. Additionally, as disorder increases, the localization length of MZMs grows (Fig.~\ref{fig5}(a)),  leading to greater overlap with the left MZM ($\gamma_1$) during the movement of the right MZM ($\gamma_2$)  from site $j=16$ to site $j=10$. This overlap contributes to a sudden increase in diabatic error for $\sigma \geq 0.4$~\cite{Thomale3, Rachel}.  In Fig.~\ref{fig5}(c) we show the $LDOS(\omega, j, t)$ for $j=10$ at $t/\tau = 6$ and $\sigma=0.2$ for a particular realization. We found sharp electron and hole peaks of equal height, confirming the movement of the MZM from $j=16$ to $j=10$ in a time $t/\tau = 6$. 

These findings highlight that quantum dot systems provide a robust platform for quantum information processing using MZMs, even in the presence of disorder, in contrast to nanowire systems~\cite{Pan}.  
For moderately sized quantum-dot systems with disorder and interactions, the diabatic error remains small during the movement of MZMs at speeds achievable in quantum-dot experiments using time-dependent gate potentials.  We also calculated the diabatic error upon introducing disorder in the on-site potential and found that it remains relatively small up to moderate disorder strengths compared to semiconducting nanowire systems (details are provided in Appendix \ref{appendix:onsite}). 
 

\subsection{Movement of the MZMs by Tuning Multiple Gates Simultaneously}\label{sec:results:gate}

\begin{figure*}
\centering
\begin{overpic}[width=2.0\columnwidth]{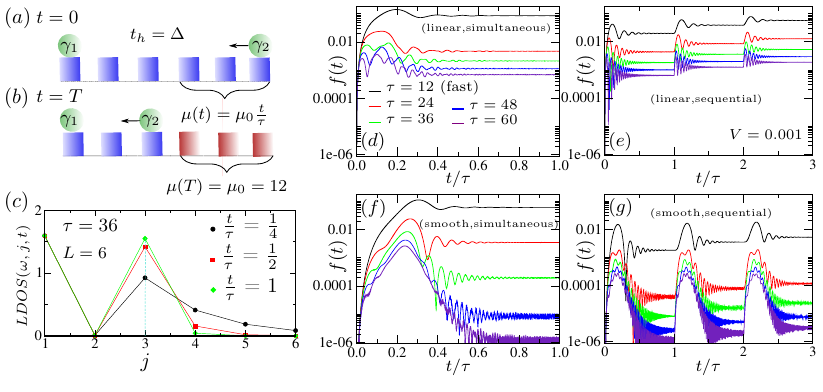}
\end{overpic}
\caption{Schematic of the movement of the right-edge MZM, $\gamma_2$, by simultaneously tuning three gates in a six-quantum-dot system. (a) At time $t = 0$, the system contains two MZMs ($\gamma_1$, $\gamma_2$), located at the left and right edges, when operating at the sweet spot $t_h = \Delta$. (b) At time $t = T$, the right-edge MZM has moved to site $j = 3$ via simultaneous tuning of the time-dependent onsite chemical potentials, $\mu(t)$, applied to individual quantum dots. (c) Time-dependent local density of states, $LDOS(\omega, j, t)$, for different values of $t/\tau$ using $\tau = 36$. (d,f) Diabatic error, $f(t)$, for the simultaneous movement of three gate potentials as a function of $t/\tau$ and for different values of $\tau$ using: (d) linear-ramp protocol, (f) sine-squared ramp protocol. (e,g) Diabatic error, $f(t)$, for the sequential movement of the three gate potentials as a function of $t/\tau$ and for different values of $\tau$ using: (e) linear-ramp protocol, (g) sine-squared ramp protocol.}
  \label{fig6}
\end{figure*}


In quantum-dot experiments, the gate attached to an individual quantum dot can be tuned either sequentially or simultaneously across multiple gates~\cite{Dvir,ChunX}. In this subsection, we compare the diabatic error during the movement of the right-edge MZM, $\gamma_2$, for two scenarios: (i) when multiple gates are tuned simultaneously, and (ii) when the gates are tuned sequentially, one by one.
To illustrate this, we consider a setup involving six quantum dots to perform the movement of the right-edge MZM, $\gamma_2$ (see Fig.~\ref{fig6}(a)), which is experimentally feasible in current quantum-dot setups. We then tune the chemical potential, $\mu(t)=\mu_0 t/\tau$ (where $\mu_0=12$), of the three rightmost quantum dots simultaneously (see Fig.~\ref{fig6}(b)) at a quench rate of $1/\tau$. We chose very small values of repulsive Coulomb interaction $V=0.001$, to prevent the overlap of MZMs  in the small system size $L=6$ that we consider for our numerical simulation.  In Fig.~\ref{fig6}(c), we present the time-dependent, site-dependent local density-of-states, $LDOS(\omega, j, t)$, for different values of $t/\tau$, when the three gate potentials, $\mu(t)$, are tuned simultaneously with $\tau = 36$. Interestingly, for smaller values of $t/\tau$, the local density-of-states exhibits delocalized peaks spanning three sites, with a higher weight on site $j = 3$. At the final time, $t/\tau = 1$, the $LDOS(\omega, j, t)$ displays two localized peaks on sites $j = 1$ and $j = 3$, indicating that the right-edge MZM, $\gamma_2$, has been successfully transferred from site $j = 6$ to $j = 3$.

In Fig.~\ref{fig6}(d), we plot the diabatic error, $f(t)$, for the movement of $\gamma_2$ under the simultaneous tuning of the chemical potential of the three quantum dots, for different values of $\tau$. For fast movement ($\tau = 12$), the diabatic error is significantly larger and exceeds the values observed for the sequential movement of the MZM [see Fig.~\ref{fig6}(d) and Fig.~\ref{fig6}(e)]. This behavior is expected, as fast tuning injects a large amount of energy over a short period, leading to increased quasi-particle excitations above the gap. Consequently, the time-evolving state, $|\psi(t)\rangle$, exhibits a substantial overlap with multiple high-energy states, resulting in a larger diabatic error. For lower values of the quench rate, $1/\tau$, the diabatic error in both cases decreases to a smaller value (see Fig.~\ref{fig6}(d)). Interestingly, the diabatic error, $f(t)$, for the simultaneous tuning of three gates is smaller (Fig.~\ref{fig6}(d)) compared to the sequential movement of the right-edge MZM (Fig.~\ref{fig6}(e)), even though the total time required for the sequential movement of $\gamma_2$ is three times longer than that for the simultaneous movement. This difference arises because, in the sequential movement, the energy levels undergo three distinct changes as the chemical potential, $\mu_i(t)$, is tuned one gate at a time during the movement of $\gamma_2$. This introduces nonanalytic behavior for each gate-tuning event and leads to an accumulation of diabatic error with increase in time~\cite{Truong,Garraway,Vitanov}. In contrast, with simultaneous multiple gate tuning (at a slower quench rate, $1/\tau$), the effective gap is modified only once, significantly reducing the occurrence of nonanalytic behavior compared to the sequential movement of $\gamma_2$.  

In Figs.~\ref{fig6}(f,g), we analyze the diabatic error using a smooth time-dependent onsite potential~\cite{Truong,Sekania}, $\mu(t) = \mu_0 \sin^2\left(\frac{\pi t}{2\tau}\right)$, for two cases: (i) when multiple gates are tuned simultaneously (Fig.~\ref{fig6}(f)) and (ii) when the gates are tuned sequentially, one by one (Fig.~\ref{fig6}(g)).  For faster movement ($\tau = 12,24$) with simultaneous tuning of the smooth time-dependent gate potential, the diabatic error is comparable to that of linear gate potential tuning. This is expected, as a large amount of energy is quenched in a short time. Interestingly, for sequential and smooth time-dependent gate tuning (Fig.~\ref{fig6}(g)), the diabatic error remains significantly lower in the range $24 \leq \tau \leq 48$ compared to all other cases of diabatic movement (Figs.~\ref{fig6}(d,e,f)). This reduction occurs because sequential smooth tuning mitigates nonanalytic behavior at each gate-tuning event in comparison to linear gate potential tuning~\cite{Garrido}.  Remarkably, for larger $\tau \geq 60$ (slower tuning), the simultaneous tuning of the smooth time-dependent gate potential (Fig.~\ref{fig6}(f)) results in the lowest diabatic error ($\sim 10^{-6}$) over time.

Next, we estimate the time scale (in SI units) required to move the right-edge MZM, $\gamma_2$, from site $j = 6$ to $j = 3$ during the multisite tuning of gate potentials with a switching time of $\tau = 36$. Using the uncertainty relation, $\tau \hbar / \Delta_{\text{eff}}$, where the effective gap in quantum dot systems is $25  \mu\text{eV} < \Delta_{\text{eff}} < 75  \mu\text{eV}$ ~\cite{Dvir,Zatelli}, we estimate the required time to be in the range of 0.316 ns to 0.947 ns. In Ref.~\cite{ChunX}, the lifetime of the target qubit was estimated to be approximately 1 ms using a five-quantum-dot system, which is significantly larger than our estimated time to move the MZM, $\gamma_2$, from site $j = 6$ to $j = 3$. For the sequential movement of the MZM, the estimated time is approximately three times longer than that for the simultaneous tuning of the three gates. 

The above results demonstrate that the movement of MZMs in quantum-dot systems for quantum information processing can be completed well within known qubit lifetimes~\cite{ChunX}, ensuring robustness against decoherence and quasiparticle poisoning during the process.  Furthermore, we conclude that an optimal strategy for quantum information processing (such as braiding or fusion) in a network of quantum dots involves a combination of simultaneous and sequential MZM movement using a smooth time-dependent gate potential. The simultaneous movement of MZMs enhances delocalization over multiple sites, whereas the sequential movement provides better control and preserves the localized nature of MZMs.


\section{Conclusion and Discussion} \label{sec:conclusion}

In conclusion, we investigate the diabatic errors associated with the movement of Majorana zero modes (MZMs) in quantum dot systems at the sweet spot, where the electronic hopping equals the superconducting coupling. Our findings highlight that the localized onsite nature of MZMs in quantum dots at the sweet spot, combined with the advantages of a moderately sized system, provides enhanced robustness and significantly reduces errors during MZM movement, in contrast to nanowire systems where MZMs exhibit localization but across multiple sites. Using time-dependent many-body numerical methods, we analyze the diabatic errors during MZM movement. Our results reveal that maintaining the well-localized behavior of MZMs during movement requires an optimal potential wall height. Importantly, probing the time-dependent local density-of-states is essential for capturing the well-localized transfer of MZMs, as fidelity measurements alone are insufficient.

In the presence of nearest-neighbor repulsive Coulomb interactions, minimizing diabatic errors requires slower quench rates due to the effective gap reduction  and increase in coherence length with  stronger interactions.  Interestingly, we find that moderate levels of disorder in hopping and pairing terms ($\sigma \leq 0.4$) do not significantly impact the diabatic errors. However, with further increase in disorder, we observe a sharp increase in diabatic errors, signifying MZM quality degradation caused by increased coherence length and reduced effective gap. We also observed a similar trends in diabatic error for the movment of MZMs, in case of onsite quench disorder.  

Our comparative study using linear and smooth gate tuning functions revealed that  
smooth gate tuning generates less diabatic error compared to linear gate potential tuning during MZM movement.  We also find that simultaneous smooth gate potential tuning outperforms sequential tuning for slower quench rates,  making it advantageous for efficient MZM transport.  
However, in quantum dot networks designed for braiding experiments, a combination of simultaneous and sequential tuning offers greater control and better preserves the localized nature of MZMs. Such controlled movements are essential for minimizing overlaps with other MZMs, ensuring optimized quantum information processing in quantum dot networks using MZMs.

Finally, using just six quantum-dots, we estimated the time scale to move the MZM  from end sites to third sites. We show that our predicted time scale (~0.3ns to 0.9ns) to move MZM in realistic setups is well within the qubit lifetime (~1ms). We strongly believe that our novel theoretical results of diabatic error for the movement of Majoranas in the one-dimensional chain  will be helpful for fusion and braiding experiments in the recently developed quantum-dot setups. 
Our predictions, based on time-dependent LDOS and fidelity measurements, align with current experimental capabilities, paving the way for future advancements in quantum-dot-based topological quantum computation.

\begin{acknowledgments}
The work of BP, SO, and ED was supported by the U.S. Department of Energy (DOE), Office of Science, Basic Energy Sciences (BES), Materials Sciences and Engineering Division. The work of G.A. was supported by the U.S. Department of Energy, Office of Science, National Quantum Information Science Research Centers, Quantum Science Center. The work of G. K. G. was supported  by the Texas Center for Superconductivity, University of Houston.
\end{acknowledgments}

{\it Note added:}
All the data and simulation code are available from the corresponding author upon reasonable request.

\appendix
\section{Non-analytic behavior of many-body gap} \label{appendix:gap}
\begin{figure}[t]
\centering
\includegraphics[width=0.48\textwidth]{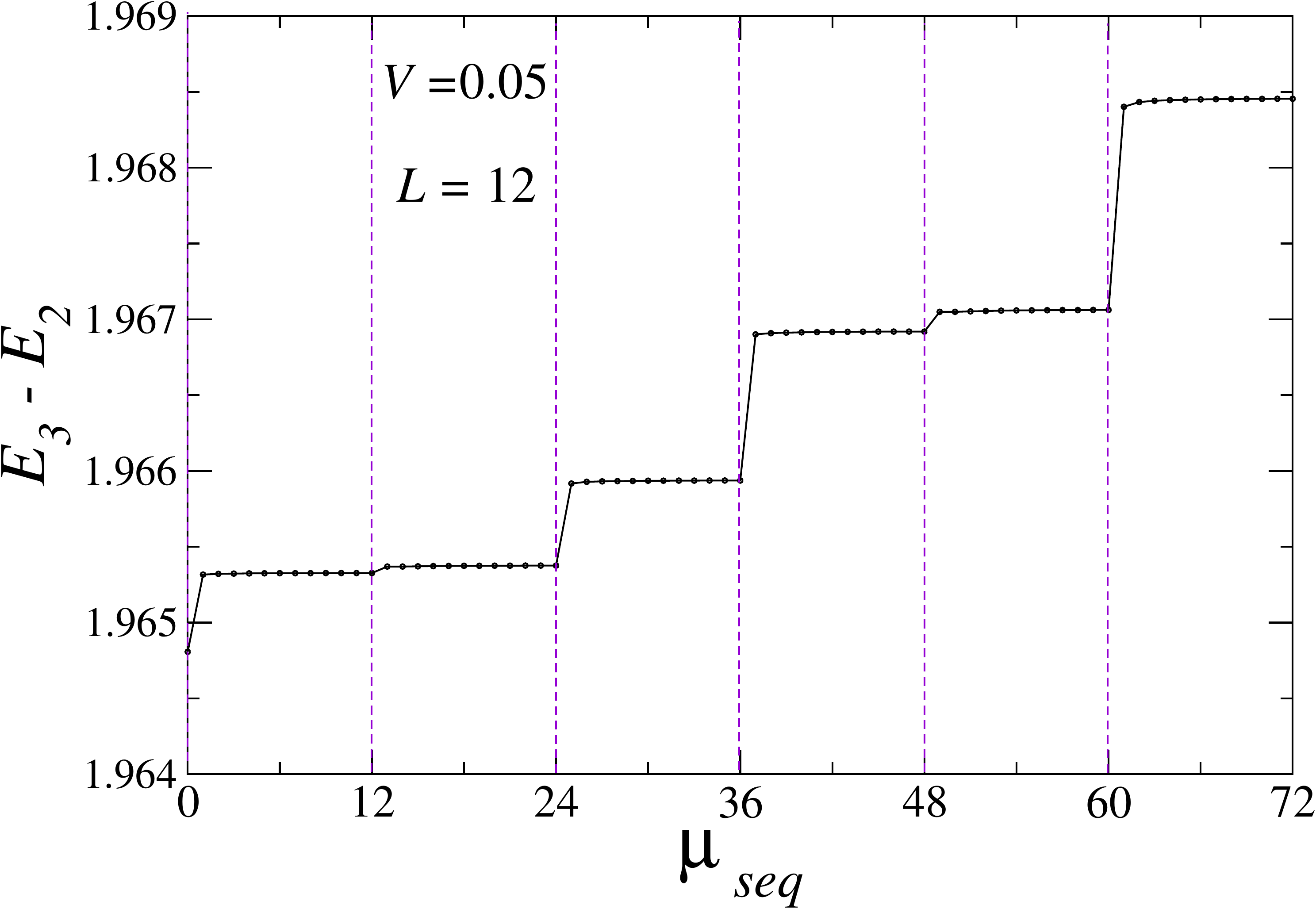}
\caption{Difference in many-body energy levels $E_3$ (first excited stated above the gap) and $E_2$ (degenerate ground state) with change in sequential onsite chemical potential $\mu_i$ (from site $j=12 \to 6$) as a parameter ($\mu_{seq} = \sum_i \mu_i$). For the numerical calculation, we use full diagonalization to find the many-body energy spectrum for $L=12$ sites and $V=0.05, \Delta=t_h=1$}   
\label{fig7}
\end{figure}

To investigate the origin of the oscillations in the adiabatic error, we examined the variation in many-body energy levels as the static onsite chemical potential changes linearly, defined as $\mu_i = \mu_0 \cdot n / 12$ (where $n$ is an integer varying from 0 to 12). In Fig.~\ref{fig7}, we present the energy difference between the first excited state ($E_3$) and the ground state ($E_2$), which is nearly degenerate with $E_1$, as a function of the sequential total gate potential. The total gate potential is defined as $\mu_{\text{seq}} = \sum_i \mu_i$, with $\mu_0 = 12$.  
As illustrated in Fig.~\ref{fig7}, the energy difference ($E_3 - E_2$) exhibits noticeable jumps or changes only at the initial stages of increasing the chemical potential ($\mu_i$) at each site.

\section{Disorder in on-site potential} \label{appendix:onsite}
\begin{figure}[t]
\centering
\includegraphics[width=0.48\textwidth]{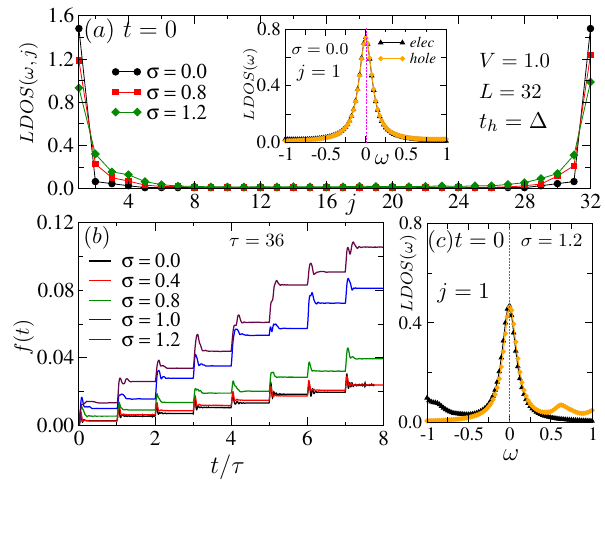}
\caption{Disorder averaged local density-of-states $LDOS(\omega,j)$ vs $j$ at $t=0$  for different                values of disorder strength $\sigma$, introduced in the onsite chemical potential $\mu_j$, and $t_h=\Delta=1$, $V=1.0$, $L=32$, $\mu_0=12$.
    Inset shows electron and hole parts of the local density-of-states $LDOS(\omega)$ vs $\omega$ at $t=0$, $j=1$ and for $\sigma=0$.
    (c) Disordered averaged electron and hole part of the local density-of-states $LDOS(\omega)$ vs $\omega$ at $t=0$, $j=1$ and for $\sigma=1.2$
   (b)  The disorder averaged diabatic error $f(t)$ in the system while moving the
    right MZM $\gamma_2$ from site $j=32 \to 24$, for different values of $\sigma$ and with increase in time $t/\tau$.
}
\label{fig8}
\end{figure}

Figure \ref{fig8} shows the effect of disorder similar to Fig. \ref{fig4} except for on-site disorder. We study the movement of the right-edge MZM, $\gamma_2$, for a moderately sized quantum dot array with $L=32$. In Fig. \ref{fig8} (a), we show the site-dependent $LDOS(\omega, j,t)$ at $t=0$ for $t_h = \Delta = 1$ and $V = 1.0$. We found MZMs still localized at both edges but they develop a tail in the bulk, which increases with the strength of disorder. In the inset, we show the $LDOS(\omega, j, t)$ at site $j=1$ and time $t = 0$. We find equal-height electron and hole peaks, confirming the presence of MZMs. In Fig \ref{fig8}(b), we show the accumulation of diabatic error with time as we move the right-edge MZM, $\gamma_2$, from site $j = 32\to 24$. For small and moderate values of disorder strength, namely for $\sigma = 0.4$ and $0.8$, the effect of disorder is small. For larger disorder strengths, namely for $\sigma  = 1.0$ and $1.2$, the diabatic error increases fast. In Fig. \ref{fig8}(c) we show the $LDOS(\omega, j, t)$ for $j=1$, $t=0$, and $\sigma = 1.2$ for a particular realization. We still find equal-height peaks for electron and hole, confirming the presence of a MZM, but the movement of these MZMs acquire large diabatic errors due to low-quality MZMs. We also see finite $LDOS$ away from $\omega = 0$ because disorder introduces localized states close to the superconducting coherence peaks. 
{{These results demonstrate that the diabatic error remains small up to moderate values of disorder and interactions. In contrast, for nanowire systems, the effective gap is expected to nearly close under moderate disorder strength and for system sizes comparable to quantum-dot systems. Meanwhile, the effective gap in quantum-dot systems remains robust under moderate disorder and interactions, resulting in lower diabatic errors compared to nanowire systems~\cite{Pan}.}}



\end{document}